\documentstyle[aps]{revtex}
\begin{document}
\draft
\title {Periodic features in the
Dynamic Structure Factor of the Quasiperiodic Period-doubling
Lattice}

\author{Anathnath Ghosh and S. N. Karmakar\cite{karm}}
\address{Saha Institute of Nuclear Physics\\
1/AF, Bidhannagar, Calcutta-700064, India.}
\maketitle

\begin{abstract}
    We present an exact real-space renormalization group (RSRG)
method for evaluating the dynamic structure factor of an infinite
one-dimensional quasiperiodic period-doubling (PD) lattice. We
observe that for every normal mode frequency of the chain, the
dynamic structure factor $S(q,\omega )$ always exhibits periodicity
with respect to the wave vector $q$ and the presence of such
periodicity even in absence of translational invariance in the
system is quite surprising. Our analysis shows that this periodicity
in $S(q,\omega )$ actually indicates the presence of delocalized
phonon modes in the PD chain. The Brillouin Zones of the lattice
are found to have a hierarchical structure and the dispersion
relation gives both the acoustic as well as optical branches. The
phonon dispersion curves have a nested structure and we have 
shown that it is actually the superposition of the dispersion
curves of an infinite set of periodic lattices.
\end{abstract}
\pacs{PACS Numbers: 71.23.Ft, 61.44.-n, 63.20.Dj }

 \section{Introduction}
The discovery of icosahedral phase in Al-Mn alloys by Shechtman
et al.\ \cite{sham} initiated experimental and theoretical
research on quasiperiodic systems. Theoretical investigations
are mostly based on one-dimensional quasiperiodic systems,
as they can be
described by simple models for pointing out the relevant 
properties of these systems.   
There are many interesting works in the literature 
concerning the electronic \cite{sok,koh,liu,chak,bov,mac,roc,igu}
and vibrational properties \cite{luck,ash,kar,sal,san,hira}
of these lattices.  At present there is considerable amount of
understanding about the nature of the eigenfunctions and the
eigenvalue spectrum for this class of deterministic hierarchical
systems lacking translational invariance. The general consensus
is that the properties like critical states, Cantor-set spectrum,
power-law behavior of the density of states etc. \cite{koh,roc}
are the characteristic
features of the quasiperiodic or aperiodic systems, though there
are evidences of extended states \cite{chak,mac}
in such systems. Moreover, there are also some reports on the
anomalous transport
\cite{roc} and optical localization \cite{igu} properties of
 these systems.
But so far very less attention has been given regarding
the dynamic response function of these lattices, a quantity
which is directly connected to the scattering cross-section
obtained from neutron diffraction or X-ray scattering experiments.
Even the computation of the dynamic structure factor for 
one-dimensional quasiperiodic system is quite meaningful, as
it is now possible to fabricate semiconducting superlattices
epitaxially \cite{mer},
 and one can perform neutron diffraction or X-ray
scattering experiments on these samples. 
Therefore, we can directly compare
the theoretical predictions with the experimental results. As
we will see, the study of the dynamic structure factor also
gives information about the nature of the eigenmodes and 
consequently one can predict the character 
of the modes from experimental observations.

In this paper we have addressed the dynamical properties of
quasiperiodic systems. We have computed the dynamical
structure factor for the quasiperiodic period-doubling (PD)
chain using a technique based on the real-space renormalization
group (RSRG) method. Here we present a formalism for calculating
the dynamic structure factor of the system using ideas developed in
Ref.\ \cite{kar}. This is an exact method  for computing the dynamic
structure factor for a class of one-dimensional quasiperiodic
chains, where the original lattice can be split 
into a finite number of self-similar sublattices.

Our study
of the dynamic structure factor also gives information about the
nature of the eigenmodes. The dispersion relation for the
PD chain can be easily obtained from the dynamic structure
factor,  and the normal mode
frequencies become readily available.  Hence it is  possible
to characterize the nature of the states  using the transfer matrix
technique. The most interesting observation  is that
for every normal mode frequency of the PD chain, the dynamic
structure factor becomes a periodic function of the wavevector,
and it actually signifies that
 this lattice supports delocalized classical vibrational
modes for all these eigenmodes. We have also shown
that for a certain choice of the parameters of the system,
 the dynamics of the PD chain exactly coincide with that
of various periodic chains having different periods, and the 
corresponding features are reflected in the dispersion relation.
Our study also reveals that one can infer about the
delocalized nature of the states from a mere inspection 
of the X-ray or neutron scattering experiment data.

In this work, though we have concentrated our attention 
on the quasiperiodic PD chain, the ideas are quite
general and will be applicable in many other quasiperiodic
systems. We have studied  the PD chain as a prototype
example for illustrating our method. An added interest
in taking the PD lattice is that the problem of the
existence of delocalized modes in this lattice can
also be studied on the basis of dimmer-type
correlation \cite{chak,wu} among the atoms.
But we would like to stress that though the method 
of Ref.\ \cite{chak} is theoretically very elegant, from the 
practical point of view its applicability is highly limited.
The procedure of Ref.\ \cite{chak} actually gives only few
delocalized normal mode frequencies, 
and the determination of the entire spectrum
becomes a formidable computational task as it requires 
the solutions of polynomial equations with progressively
higher and higher degree. On the other hand, the dynamic
structure factor automatically gives the entire spectrum 
and the nature of the eigenmode corresponding to every 
normal mode frequency can be easily characterized.
An interesting feature of the PD chain is that 
the trace map associated with this lattice leads to
a polynomial invariant which should normally ensure
the critical nature
of all the eigenstates as well as a Cantor-set spectrum
\cite{koh,luck}.
However, in this work we find that this lattice supports 
an infinity of delocalized states in spite of the Cantor-set
nature of the spectrum.

This paper has been organized as follows. We first describe
our model in Section II. We present in Section III the RSRG
scheme for finding the dynamic structure factor of a quasiperiodic
system, while we briefly outline the transfer matrix method
in the next Section. Section V is devoted for the analysis
of our results and we conclude in Section VI.

\section{Description of the model}
We describe the vibrational properties of the PD chain
considering nearest neighbor harmonic interaction 
among the atoms. A portion of the PD chain is shown in Fig.\ 1. The
PD sequence can be generated from two symbols $L$ and $S$
using recursively the substitution rule $L \rightarrow LS$
and $S \rightarrow LL$. Thus $L$, $LS$, $LSLL$, $LSLLLSLS$, \ldots etc.
are the first few generations of the PD sequence. Now we can
construct a PD chain by considering the symbols $L$ and $S$
as representing `long' ($L$) and `short' ($S$) bonds in the chain.
In our model  $m_\alpha$,
$m_\beta$ and $m_\gamma$ are the masses of the atoms flanked
by $L-L$, $L-S$ and $S-L$ bonds respectively, and, $k_L$ and
$k_S$ are respectively the spring constants across the long
and short bonds. It should be noted that this kind of 
labeling of the sites and the bonds is essential for implementing
 the RSRG decimation procedure of Southern et al.\ \cite{sou}.
We shall see that successive renormalized lattices 
can be represented in terms of only these five parameters
and the number of the parameters do not grow with iteration.
In other words, within this labeling scheme, the form
of the equations of motion of the system remains invariant
under renormalization. The well-known on-site model
and the bond-model for the PD chain turn out to be 
the special cases of this general situation. 

The expression for the dynamic structure factor $S(q,\omega)$
corresponding to the wavevector $q$ and frequency $\omega$
can be written in terms of the
single particle Green's functions as follows

\begin{equation}
S(q,\omega) =
\lim_{\delta \rightarrow 0^+} \lim_{N \rightarrow \infty} 
\mbox{Im} (G_N(q,\omega +i\delta))~,
\label{eq:m1}
\end{equation}
where $G_N(q,\omega) = (1/N) \sum_{ij} e^{iq (r_i - r_j)}  G_{ij} $ .
Here $r_i$ is the position vector of the $i$th atom and $N$ represents
the total number of atoms in the chain. In the harmonic approximation
with nearest neighbor coupling between the atoms, the single particle
Green's functions $G_{ij}$'s  satisfy the following equations
of motion

\begin{equation}
\epsilon_i G_{ij} =
-\delta_{ij} + k_{i,i+1} G_{i+1j} + k_{i,i-1} G_{i-1j~,}
\label{eq:m2}
\end{equation}
where $\epsilon_i =  k_{i,i+1} + k_{i,i-1} - m_i\omega^2$,  $m_i$
being the mass of the $i$th atom, and $k_{ij}$ the spring constant
between the  $i$th and $j$th atoms.

Now the calculation of $S(q,\omega)$ for a translationally 
invariant system is very straight forward. $G_N(q,\omega)$
can be easily computed from a Fourier transformed version of
the Eqs.\ (\ref{eq:m2}). However, for quasiperiodic
systems the determination of the sum in $G_N(q,\omega)$
is not trivial. We shall show in the next section that this
sum can be evaluated in a very elegant way using
RSRG method. The calculation of $S(q,\omega)$ essentially
amounts to the iteration of certain recursion relations and it
can be computed with arbitrarily high accuracy.

\section{RSRG Scheme for the Determination of $S(q,\omega)$}

Let us now consider the sum $G^i(q,\omega) = 
 \sum_j  e^{iq (r_i - r_j)} G_{ij}$.
This sum is independent of the index $i$ in a periodic system
due to the translational  invariance. However, in the case of a
quasiperiodic system, it depends on the index $i$
as no two sites are equivalent in the quasiperiodic lattice and
 from Eqs.\ (\ref{eq:m2})
we obtain

\begin{equation}
\epsilon_i G^i(q,\omega) = 
-F_i + k_{i,i+1} e^{-i q (r_{i+1} - r_i)} G^{i+1}(q,\omega) +
k_{i,i-1} e^{i q (r_i - r_{i-1})} G^{i-1}(q,\omega)~,
\label{eq:m3}
\end{equation}
where all $F_i$'s are initially equal to unity. The use of new
notations $F_i$'s for representing unity does not directly follow
from Eqs.\ (\ref{eq:m2}). We have introduced these notations 
by hand and we will see how it facilitates the determination
of $S(q,\omega)$ in quasiperiodic lattices. Even though 
all $F_i$'s are initially the same, they  become different 
upon renormalization. However, from the symmetry of
the lattice, we observe that there will be only 
three distinct type of $F_i$'s, and we can identify them
as $F_\alpha$, $F_\beta$ and $F_\gamma$ corresponding
to the $\alpha$, $\beta$ and $\gamma$ site of the lattice.

It is now necessary to determine all $G^i(q,\omega)$'s from 
Eqs.\ (\ref{eq:m3}) for evaluating $G_N(q,\omega)$ since we have

\begin{equation}
G_N(q,\omega) = 
(1/N) \sum_i G^i(q,\omega)~.
\label{eq:m4}
\end{equation}
In the quasiperiodic PD chain $G^i(q,\omega)$'s are all distinct
and instead of determining them directly from Eqs.\ (\ref{eq:m3}),
we use RSRG technique
for finding them. We split the
original chain into two self-similar sublattices $ \Omega$ and
$\Gamma$ as shown in Fig.\ 1. The $\Omega$-sublattice is obtained
by eliminating the sites using the decimation rules
$LS \rightarrow L$ and $ LL \rightarrow S$, while the
corresponding rules for the $\Gamma$-sublattice
are $SL \rightarrow L$ and $LL \rightarrow S$. Both $\Omega$
and $\Gamma$ sublattices again form two new PD chains
at some inflated length scale. All the sites of the original
PD lattice are distributed among these two sublattices, 
and thus they are complementary to each
other (see Fig. 1). This complementary nature of $\Omega$ and
$\Gamma$ sublattices also ensures that no information is lost
by introducing this splitting procedure.
It is now possible to generate
two sets of renormalized equations for $G^i(q,\omega)$'s, one
corresponding to $\Omega$-sublattice while the other for the 
$\Gamma$-sublattice. The equations for $\Omega$-sublattice
can be obtained from Eqs.\ (\ref{eq:m3}) by eliminating
 all $G^i(q,\omega)$'s belonging to the $\Gamma$-sublattice, 
and the resulting equations
can be cast in the same form as that of the original set 
of equations (\ref{eq:m3}), provided we rename the sites appropriately 
 and renormalize the 
parameters as follows 

\begin{eqnarray}
\epsilon_\alpha^{'}  &=&
\epsilon_\gamma - \omega_\beta (k_L^2 + k_S^2), \nonumber \\ 
\epsilon_\beta^{'} &=& 
\epsilon_\gamma - (k_L^2 \omega_\alpha + k_S^2 \omega_\beta), \nonumber \\
\epsilon_\gamma^{'} &=& 
\epsilon_\alpha - k_L^2 (\omega_\alpha + \omega_\beta), \nonumber \\
F_\alpha^{'} &=& 
F_\gamma +F_\beta k_S \omega_\beta e^{i q a_S} +
F_\beta k_L \omega_\beta e^{-i q a_L}, \nonumber \\ 
F_\beta^{'} &=& 
F_\gamma +F_\alpha k_L \omega_\alpha e^{-i q a_L} +
F_\beta k_S \omega_\beta e^{i q a_S}, \nonumber \\ 
F_\gamma^{'} &=& 
F_\alpha +F_\alpha k_L \omega_\alpha e^{i q a_L} +
F_\beta k_L \omega_\beta e^{-i q a_L}, \nonumber \\ 
k_L^{'} &=& k_L k_S \omega_\beta~, \nonumber \\ 
k_S^{'} &=& k_L^2 \omega_\alpha~, \nonumber \\
a_L^{'} &=& a_L + a_S~, \nonumber \\
a_S^{'} &=& a_L + a_L~.
\label{eq:a}
\end{eqnarray}
Similarly for the $\Gamma$-sublattice the recursion relations can
be  written as
\begin{eqnarray}
\epsilon_\alpha^{'}  &=&
\epsilon_\beta - \omega_\gamma (k_L^2 + k_S^2), \nonumber \\ 
\epsilon_\beta^{'} &=& 
\epsilon_\alpha - k_L^2 (\omega_\alpha + \omega_\gamma), \nonumber \\
\epsilon_\gamma^{'} &=& 
\epsilon_\beta - (k_S^2 \omega_\gamma + k_L^2 \omega_\alpha), \nonumber \\
F_\alpha^{'} &=& 
F_\beta +F_\gamma k_L \omega_\gamma e^{i q a_L} +
F_\gamma k_S \omega_\gamma e^{-i q a_S}, \nonumber \\ 
F_\beta^{'} &=& 
F_\alpha +F_\beta k_L \omega_\alpha e^{-i q a_L} +
F_\gamma k_S \omega_\gamma e^{i q a_S}, \nonumber \\ 
F_\gamma^{'} &=& 
F_\beta +F_\alpha k_L \omega_\alpha e^{i q a_L} +
F_\gamma k_S \omega_\gamma e^{-i q a_S}, \nonumber \\ 
k_L^{'} &=& k_L k_S \omega_\gamma~, \nonumber \\ 
k_S^{'} &=& k_L^2 \omega_\alpha~, \nonumber \\
a_L^{'} &=& a_L + a_S~, \nonumber \\
a_S^{'} &=& a_L + a_L~.
\label{eq:b}
\end{eqnarray} 
Here $\epsilon_i = k_{i,i+1} + k_{i,i-1} - m_i \omega^2$,
 $\omega_i = 1/\epsilon_i$ and  $i$ refers to
$\alpha$, $\beta$ or $\gamma$. In Eqs. (\ref{eq:a}) and
(\ref{eq:b}) $a_L$ and $a_S$ respectively represent the
`long' and `short' bond lengths.
Let us represent the above two transformations  by $T_{\Omega(\Gamma)}$ and
denote the renormalized Green's functions as $G_{\Omega(\Gamma)}^i$. 
Now we can recast $G_N(q,\omega)$ as

\begin{equation}
G_N(q,\omega) = 
p_{(\Omega)} G_N^{(\Omega)}(q,\omega) +p_{(\Gamma)} G_N^{(\Gamma)}(q,\omega),
\label{eq:m5}
\end{equation}
where $G_N^{(\lambda)}(q,\omega)  = (1/N_{\lambda}) \sum_{i \in \lambda}
G_{(\lambda)}^i(q,\omega)$. Here $\lambda$ can be either
$\Omega$ or $\Gamma$, and $N_{\Omega (\Gamma)}$ is the number
of sites in the renormalized $\Omega$$(\Gamma)$ sublattice. The
coefficients $p_{(\Omega)}$ and $p_{(\Gamma)}$ denote the fractions of
total sites which belong to $\Omega$ and $\Gamma$ sublattices
respectively, i.e., $p_{(\Omega)}=N_\Omega /N$ and
$p_{(\Gamma)}=N_\Gamma /N$. The expressions
for $G_N(q,\omega)$ and $G_N^{(\lambda)}(q,\omega)$ are
structurally same, the former being defined in terms 
of the parameters of the original PD lattice while the latter 
refers to those of the renormalized $\lambda$-sublattice. 
Eq.\ (\ref{eq:m5}) 
shows that $G_N(q,\omega)$ is a linear combination of
$G_N^{(\Omega)}(q,\omega)$ and $G_N^{(\Gamma)}(q,\omega)$ 
with coefficients $p_{(\Omega)}$ and $p_{(\Gamma)}$, where
$p_{(\Omega)} = p_{(\Gamma)} = 1/2$. Since each of these
sublattices again forms a new PD chain, we can treat them 
at the same footing as the original chain. Thus in the renormalized
$\Omega(\Gamma)$ sublattice, $G_N^{\Omega(\Gamma)}$ takes
the same role as that of $G_N(q,\omega)$ for the original lattice.
As we can split further each of these new chains into
$\Omega$ and $\Gamma$ sub-sublattices, it is again possible to
express both $G_N^{(\Omega)}(q,\omega)$ and $G_N^{(\Gamma)}(q,\omega)$
in the form of Eq.\ (\ref{eq:m5}). Thus we have

\begin{equation}
G_N(q,\omega) = 
p_{(\Omega \Omega)} G_N^{(\Omega \Omega)}(q,\omega) +
p_{(\Omega \Gamma)} G_N^{(\Omega \Gamma)}(q,\omega) +
p_{(\Gamma \Omega) } G_N^{(\Gamma \Omega)}(q,\omega) +
p_{(\Gamma \Gamma)} G_N^{(\Gamma \Gamma)}(q,\omega).
\label{eq:m6}
\end{equation} 
Here we denote the two branches resulting from $\Omega$ sublattice
as $(\Omega \Omega, \Omega \Gamma )$, while those from $\Gamma$-sublattice
as $(\Gamma \Omega , \Gamma \Gamma)$ . The coefficients
can be written as $p_{(\mu \nu)} = p_{(\mu)} p_{(\nu)}$, with
$\mu, \nu = \Omega $ or $\Gamma$.

If we continue the splitting procedure it will give rise to
a tree like structure and it is possible to label each sublattice
by its path in this tree. In other words,
we label a sublattice by specifying the sequence of
$\Omega$ and $\Gamma$ branches that constitute the path 
leading to the sublattice. The idea of above labeling comes from
the fact that this branching process actually gives
a family classification \cite{kar} for the sites of the PD chain,
 and we can consider this tree as the 
genealogical tree for this lattice. So finally
we can write $G_N(q,\omega)$ as

\begin{equation}
G_N(q,\omega) = 
\sum_{all~paths} p_{(path)} G_N^{(path)}(q,\omega),
\label{eq:m7}
\end{equation} 
where the sum is over all possible paths in the genealogical
tree for a  given number of branching. We terminate
each path in the genealogical tree using the criteria
that the corresponding  renormalized coupling 
constants $k_L$ and $k_S$ flow to zero at this 
stage of iteration. In this limit the  computation
of each term in the 
Eq.\ (\ref{eq:m7}) becomes trivial (see Eq.\ (\ref{eq:m3})) 
and one can express every $G_N^{(path)}$ into the
following general form

\begin{equation}
G_N^{(path)} = -\left(
 x_\alpha F_\alpha^*/\epsilon_\alpha^* +x_\beta F_\beta^*/
\epsilon_\beta^* +x_\gamma F_\gamma^*/\epsilon_\gamma^* \right)~,
\label{eq:m8}
\end{equation}
where $\epsilon_i^*$'s  and $ F_i^*$'s represent the appropriate
renormalized parameters, and $x_\alpha, x_\beta $ and
$x_\gamma$ are the concentrations of $\alpha , \beta $ and
$\gamma$ sites in the PD chain. From $G_N(q,\omega)$ 
the dynamic structure factor can be 
obtained from Eq.\ (\ref{eq:m1}).

The merit of this scheme is that one has to generate
all possible paths in the genealogical tree using a
simple algorithm, then iterate the recursion relations
$T_{\Omega}$ or $T_{\Gamma}$  sequentially along these paths, 
and finally determine $S(q,\omega)$ using 
equations (\ref{eq:m1}), (\ref{eq:m7}) and (\ref{eq:m8}).
There is no approximation involved in this
method  and the dynamic structure factor for quasiperiodic 
lattices can be obtained with arbitrary accuracy, the
accuracy level being set by the smallness of the 
renormalized values of $k_L$ and $k_S$ .

The dispersion relations for the system can be easily obtained 
from $S(q,\omega)$ using the fact that the $\omega$ and $q$
values corresponding to every non-zero
values of $S(q,\omega)$ constitutes  a point in the 
dispersion curve. Thus scanning the entire
$\omega - q$ plane for non-zero $S(q,\omega)$ 
one can get the dispersion curve for the PD chain,
and this gives the whole spectrum of normal mode frequencies 
for the system.

\section{Lattice modes and the transfer Matrix method}
In this section we  describe briefly the transfer
matrix method for determining the normal mode vibrational
states in a one-dimensional lattice. The equations of motion 
for classical vibration of the chain are

\begin{equation}
m_i \omega^2 u_i = 
k_{i,i+1} (u_i - u_{i+1}) +k_{i,i-1}(u_i - u_{i-1}),
\label{eq:m9}
\end{equation}
where $u_i$ is the displacement of the $i$th atom. In the transfer
matrix method one can cast the above equation in the following
matrix form,

\begin{equation}
\left (\begin{array}{c}
u_{i+1} \\
u_i
\end{array} \right) = 
\left (\begin{array}{cc} 
\epsilon_i /k_{i,i+1} & - k_{i,i-1}/k_{i,i+1} \\
1 & 0
\end{array} \right)
\left (\begin{array}{c}
u_i \\
u_{i-1}
\end{array} \right)
\equiv
M_i \left (\begin{array}{c}
u_i \\
u_{i-1}
\end{array} \right).
\label{eq:m10}
\end{equation}
In any normal mode, the displacement $u_n$ of any arbitrary
atom  $n$ can be obtained from the product of a series of
transfer matrices $M_{n-1} M_{n-2} \cdots M_2 M_1 $ by
specifying the initial displacements $u_1$ and $u_0$.
However, in this technique the determination  of the
amplitudes at various sites requires a prior 
knowledge of the normal mode frequency. Since
the study of the structure factor gives the entire
normal mode frequency spectrum, we can characterize
the localized or delocalized nature of the
lattice modes in quasiperiodic systems from the 
asymptotic behavior of the amplitudes at large
distances. 

At this stage, it is worthwhile to mention
that in the PD chain $\alpha $-type sites always
occur in pairs, and an analysis of this dimmer-type
correlation provides an understanding about the extended
nature of the electronic eigenfunctions in quasiperiodic and 
disordered systems \cite{chak,wu}. In a recent work, Dominguez-Adame
et al.\ \cite{san} 
have shown that this dimmer-type correlation determines
the special frequency at which delocalized vibrational
mode can exist even in a disordered one-dimensional
chain. For a PD chain there are three types of
transfer matrices $M_\alpha, M_\beta$ and $M_\gamma$ corresponding
to the $\alpha , \beta $ and $ \gamma$  sites 
in the lattice and 
it is easy to show that for
$\epsilon_\alpha = 0 $, apart from a constant
phase factor, the total transfer matrix across the
PD chain simply reduces to that of a perfectly
periodic chain with alternating $\beta$ and
$\gamma$ atoms. Consequently the system
supports delocalized vibrational mode for the
frequency $\omega$ which satisfies the equation $\epsilon_\alpha = 0$.
Since this $\alpha - \alpha$ pairing is present
in all the subsequent renormalized lattices,
in principle one can determine the full frequency 
spectrum for delocalized modes by solving 
the equations $ \epsilon_\alpha = \epsilon_\alpha^{'} = 
\epsilon_\alpha ^{''}
\cdots = 0 $. However, as pointed out earlier, the
renormalized $\epsilon_\alpha$'s are polynomials
in $\omega$, and the degree of the polynomial 
increases very rapidly with the progress of
iteration. So it becomes practically impossible
to solve these high degree equations with adequate
accuracy. Thus dimmer approach turns out to be 
inefficient for obtaining the frequency 
spectrum of the delocalized modes. On the
other hand, the dynamic structure factor gives the entire
spectrum of normal mode frequencies without much effort
and we can easily determine the nature of the normal modes using
transfer matrix method.

\section{Results and discussions}
With the above background we shall now present
in this section the results for some specific cases 
of the PD chain. We begin by considering the
on-site model of the PD chain. In the on-site model 
two types of masses
$m_A $ and $m_B$, connected by identical spring constants,
are distributed on a lattice following
PD ordering, 
and the spacing between the atoms are equal.
We can recover the on-site model by setting $m_\alpha =
m_\gamma = m_A$, $m_\beta = m_B$,  $k_L = k_S = k$
and $a_L = a_S =a$. For numerical calculations,
we choose $m_A = 2 , m_B = 1 , k = 1$ and $a = 1 $.
In the first column of Fig.\ 2, we have plotted
the dynamic structure factor $S(q,\omega)$ as a function
of $q$ for three normal mode frequencies, while the
amplitudes of the atoms in these normal modes are
shown in the second column of the figure. The first, second and third
rows of Fig.\ 2 respectively  correspond to the 
frequencies $\omega_1^2 = 1,
\omega_2^2 = 0.38196$ and $ \omega_3^2 = 1.54739$
and these frequencies respectively
satisfy the conditions $ \epsilon_\alpha  = 0$, $\epsilon_\alpha^{'} = 0$
and $ \epsilon_\alpha^{''} = 0 $. Though these normal mode
frequencies can be obtained directly from the 
dynamic structure factor itself, here we have
determined them from the above conditions for getting better
numerical accuracy.
This figure clearly depicts that for the on-site model, the
dynamic structure factor $S(q,\omega)$ becomes a 
periodic function in wavevector $q$. From Fig.\ 2a,
2c and 2e we see that the peaks in $S(q,\omega)$
corresponding to the frequencies $\omega_1$, $\omega_2$ and $\omega_3$
occur respectively at $q = (2 n +1)\pi/2$, $(2 n +1 )\pi/4$
and $(2 n +1)\pi/8$, where $n = 0,1,2 \cdots$. Also
the corresponding normal modes display periodicity
of periods  4, 8  and 16 in units of the 
lattice spacing.

Now we shall discuss quite elaborately the underlying physical
reason for this periodicity in the dynamic structure
factor of the PD chain. For this purpose we restrict 
ourselves only to the frequency $\omega = \omega_1 $.
The  frequency $\omega_1 $ is 
a root of the equation $\epsilon_\alpha = 0 $, and 
as mentioned earlier, for this frequency one
can effectively map the PD chain into an ordered
binary chain composed of alternating $\beta $ and
$\gamma$ type  atoms. The dynamic structure factor
and the dispersion relation for this periodic lattice are
respectively given by

\begin{equation}
S(q,\omega) =- \mbox{Im} \left [ \frac{\epsilon_\beta +\epsilon_\gamma
+2(k_L +k_S)\cos qa}{2(\epsilon_\beta \epsilon_\gamma -
k_L^2 - k_S^2 - 2 k_L k_S \cos 2qa )} \right ]
\label{eq:m11}
\end{equation}
and
\begin{eqnarray} 
\omega^2 & = &(1/2 m_\beta m_\gamma)   \{ (k_L + k_S )(m_\beta +
m_\gamma) \pm
[(k_L + k_S)^2(m_\beta - m_\gamma)^2 \nonumber \\  
 & + & 4 m_\beta m_\gamma (k_L^2 +k_S^2+2 k_L k_S \cos 2 q a)]^{1/2}\}.
\label{eq:m12}
\end{eqnarray}
Setting $\omega = \omega_1$, $m_\beta = 1$, $m_\gamma = 
2 $ and $k_L/k_S = 1$
 in Eq.\ (\ref{eq:m11}), in order to afford a comparison with
the above results presented in Fig.\ 2, we see that the 
positions of the peaks in $S(q,\omega)$ for this periodic
binary chain are situated at $q = (2 n +1)\pi/2$. These
positions of the peaks are exactly the same as those
observed in Fig.\ 2a for the on-site model of the chain, though
their relative intensities may differ. In other words,
at $\omega = \omega_1$, the the peaks of $S(q,\omega)$ for
the PD chain exactly coincides with that of a periodic 
lattice whose unit cell contains two atoms $\beta$ and $\gamma$.
From this correspondence we can say that the quasiperiodic
PD chain supports delocalized vibrational mode for this
frequency and the oscillation of atomic displacements
in Fig.\ 2b actually reflects this extended nature
of the mode. In a similar fashion the periodicity
of $S(q,\omega)$ in Fig.\ 2c and Fig.\ 2e can be explained,
and we see that for these cases the dynamic structure
factor resembles that of two other periodic 
lattices whose unit cells consist of four and eight
atoms respectively. Thus as far as the dynamics are concerned,
quasiperiodic PD chain effectively behaves like an
assembly of periodic lattices with periods 4, 8, $\cdots$ etc.
and it is quite natural that the corresponding 
normal modes are extended. In fact, we have observed 
that for the present choice of the parameters of the system,
every normal mode of the PD chain exactly becomes 
identical to that of an appropriate periodic lattice.

The above equivalence between the PD chain and a set of periodic
lattices become much more apparent from the dispersion
relation for the PD chain. We have determined the
dispersion relation of the PD chain by scanning the 
entire $\omega - q$ plane for non-zero values of
$S(q,\omega)$  and it has been displayed in Fig.\ 3.
In this figure we have also plotted the dispersion 
curve (solid lines) of the periodic lattice
consisting of two type of atoms $\beta$ and $\gamma$.
In the periodic case with alternating $\beta$ and $\gamma$
type atoms, we have one acoustic and one optical branch.
Fig.\ 3 shows that there is an acoustic branch in the
PD chain which coincides with the acoustic branch 
of this particular periodic lattice, and both of these
lattices have a common Brillouin Zone boundary 
at $q = \pi/2$. This indicates that there is an
acoustic branch of the PD chain which comes from
an effective periodic lattice of unit cell size two
in terms of the lattice spacing. However, in the
case of the PD chain, we observe that there are an infinite
number of acoustic branches each having its own characteristic
Brillouin Zone, and Fig.\ 3 clearly shows that the edges
of these Brillouin Zones are situated at $q = \pi/4$,
$\pi/8$,  $\cdots$ etc. In Fig. 3 we have displayed the
dispersion curve in the extended Brillouin Zone scheme and
Zone boundaries other than those mentioned above correspond
to the repetition of the various basic Brillouin Zones.
The overlap of
the various acoustic branches implies that these
Brillouin Zones have a nested structure. Thus it is 
quite interesting  that the quasiperiodic
PD chain plays the role of an infinite number
of periodic lattices whose unit cell sizes
increase as $2^n$, where $n = 1,2, \cdots \infty$.
The optical branches of the PD chain are in fact 
the superposition of all the optical branches
of these periodic lattices and naturally
it would be quite complex.
From our calculations we really observe that the optical 
branches have a very complex structure (see Fig.\ 3)
and actually there can be infinite number of the optical
branches for the PD lattice. We observe the following
additional features in the optical branch of the PD
lattice. For PD lattice we find a well-defined structure
in the region of the $\omega - q$ plane corresponding to
the gap of the periodic $\beta - \gamma$ binary chain,
which is characteristic of the quasiperiodic order of
the chain. Fig. 3 also shows that such a structure is 
absent for low values of $q$. This is due to the fact 
that in this long wave length limit (wavelength 
$\lambda >> a_L$ or $a_S$) the detailed 
structure of the lattice is not very important, and
both the periodic $\beta - \gamma$ chain and the quasiperiodic
PD chain exhibit similar behavior.

In Fig.\ 4 we have displayed $S(q,\omega )$ as a function
of $q$ for other models of the PD chain including some
additional graphs corresponding to the on-site model.
The graphs for the on-site model are shown in Fig.\ 4a 
and  4b corresponding to the frequencies 
$\omega^2 = 2.26829$
and $\omega^2 =  1.65555$ respectively. The
graphs for the bond-model are shown in Fig.\ 4c and \ 4d which
has been realized by setting $m_\alpha =m_\beta =m_\gamma$,
$k_L \ne k_S$ and $a_L \ne a_S$. In this case we choose
$m_\alpha =m_\beta =m_\gamma =1$, $k_L =1$, $k_S =2$,
$a_L =2$ and $a_S=1$ and
the  Fig.\ 4c and 4d respectively correspond to the frequencies
$\omega^2 = 2.0 $ and $\omega^2 = 3.09447$.
In Fig.\ 4e
and  4f  we have also
presented the results for the most general case of 
the PD chain and here the parameters are taken as
$m_\alpha = 1$, $m_\beta = 2$, $m_\gamma = 3$ (all
in arbitrary units), $k_S/k_L = 2$ and $a_L/a_S = 2$.
The graphs in Fig.\ 4e and 4f are plotted
for frequencies $\omega^2 = 2.0$ and $\omega^2 = 2.42427$
respectively. The most striking feature is that in every
case we observe that $S(q,\omega )$ is always periodic in
wave vector $q$.  So we can
say that this periodicity is a characteristic 
feature of the PD lattice and it is quite independent
of the specific choice of the parameters of the system.

We have demonstrated the underlying physical reason for this periodicity   
in $S(q,\omega)$ only for the on-site model of the PD 
chain, which owes its origin to the fact that 
for every normal mode frequency one can exactly 
map the system to various periodic lattices.
Similar kind of analysis is also possible in 
other cases, the unit cell structure being quite  
complex in these situations. The most interesting 
consequence of the periodicity of $S(q,\omega)$
with respect to $q$ is that the system supports
delocalized vibrational modes even in absence
of the translational invariance.  

At this stage we would like to make the following 
remark. It should be noted that there exists
a polynomial invariant $I$ associated with
the recursion relations Eq.\ (\ref{eq:a}) and (\ref{eq:b}) for the
PD chain which is given by \cite{chak} 

\begin{equation}
I= ((\epsilon_\alpha - \epsilon_\gamma) (\epsilon_\alpha -
\epsilon_\beta) - k_S^2 - k_L^2)/(2 k_L k_S) + 1.
\end{equation}
Depending upon the choice of parameters of the system,
 $I$ may be independent of $\omega$ as well as 
it may be a function of $\omega$. In the electronic case, 
one  normally 
 equates the invariant to zero for finding the
energy eigenvalues of the extended states of quasiperiodic
systems, provided those energies are allowed ones.
Similarly we can obtain
the delocalized normal mode frequencies   
from the condition $ I = 0 $. However, we have to check
whether the frequencies obtained from the above condition
give allowed eigenmodes or not.
For $m_\alpha = 1 , m_\beta = 2 ,
m_\gamma = 3 , k_S/k_L = 2 $ and $ a_L/a_S = 2$
it is found that the condition $ I = 0 $
is satisfied for $\omega^2 = 1.5$. But this is not
an allowed frequency of the system (see Ref. \cite{san}),
which is also confirmed from the calculation of $S(q,\omega)$
and it turns out that it 
is  always zero for every $q$ in this case.
 However, depending on the choice of the
parameters of the system, the condition $I=0$ may give
a few allowed delocalized normal mode frequencies. This
condition actually corresponds to the commutation of the
transfer matrices for two successive generations of the
chain, and thus leads to delocalized states. 
It should be noted that  the commutation
conditions for various successive generation always give
the same set of frequencies due to the presence of the
invariant in this system (see
Ref. \cite{mac} for details). Our RG analysis, however,
 shows that
an infinite number of delocalized eigenmodes are possible
in the PD chain due to $\alpha - \alpha$ pairing at all
length scales.

\section{Conclusions}
In conclusion, we should stress that  
 the renormalization group
method is the right tool for taking into account
the symmetry of quasiperiodic systems and the present 
RSRG scheme offers a very efficient method
for computing the dynamic structure factor 
of these systems. The merit of  the present
scheme is that it provides an exact method for the
determination
of $S(q,\omega)$ of infinite self-similar
quasiperiodic lattices, and in practice, one can 
calculate $S(q,\omega)$ with arbitrary accuracy.
This method essentially
involves the iteration of certain
recursion relations, and so it enormously reduces
the computational task. We have applied this method 
to the quasiperiodic PD chain and obtained several new
results. Apart from computing the dynamic 
structure factor, we have also determined the 
dispersion relation of this lattice. The most 
interesting result is the observation of periodicity 
in $S(q,\omega)$ with respect to $q$ for every
normal mode frequency, and this periodicity
actually gives the signature for the presence
of normal mode vibrations which are delocalized 
over the entire lattice. A direct consequence 
of the periodicity of $S(q,\omega)$ is that 
one can infer about the existence of the
delocalized vibrational modes in the system simply
by looking for the periodicity in the scattering
intensity data from diffraction experiment 
measurements.
Another important outcome of the present work 
is that the entire spectrum of frequencies 
for delocalized modes can be obtained very easily 
from the dynamic structure factor, whereas the
determination of the full
spectrum by the method of Ref.\ \cite{san} is practically
impossible.

\section*{acknowledgments}

We  thank 
Dr. T. K. Chini for helping us to prepare a diagram.

\begin{figure}
\caption{ A section of the period-doubling chain illustrating
the sublattice splitting.
 The symbols $\bullet$, $\triangle$
and $\bigcirc$ respectively represent $\alpha$, $\beta$ and
$\gamma$ sites.}

\caption{Plot of $S(q,\omega)$ versus  $q$,
and $u_n $ versus $n$ for the on-site model with
$m_\alpha = 
m_\gamma = 2, m_\beta = 1 $ and $k_L / k_S =1, a_L / a_S =1 $. 
First, second and third rows respectively correspond to
 $ \omega^2 = 1.0$,  
 $\omega^2  = 0.38196$ and
 $ \omega^2 = 1.54739$.}

\caption{The dispersion relation for the period-doubling lattice 
with the same parameters as those in Fig. 2.}

\caption{Plot of $S(q,\omega)$ versus $q$. On-site model
with parameters $m_\alpha =m_\gamma =2$, $m_\beta =1$,
$k_L/k_S=1$ and $a_L/a_S=1$ for (a) $\omega^2=2.26829$
and (b) $\omega^2=1.65555$. Bond-model with parameters
$m_\alpha =m_\beta =m_\gamma =1$, $k_S/k_L=2$ and $a_L/a_S=2$
for (c) $\omega^2=2$ and (d) $\omega^2=3.09447$. Mixed model
with parameters $m_\alpha =1$, $m_\beta=2$, $m_\gamma =3$,
$k_S/k_L=2$ and $a_L/a_S=2$ for (e) $\omega^2=2$ and (f)
$\omega^2 =2.42427$.}
\end{figure}
\newpage
\pagestyle{empty}
\setlength{\unitlength}{0.04in}
%\begin{center}
%\begin{picture}(50,30)(0,0)
\begin{picture}(140,40)(-20,-20)
\thicklines
\put(-10,0){\circle*{2}}
%\put(-11,25){\makebox(0,0){{\tiny -10}}}
\put(-10,0){\dashbox{.5}(0,20)}
\put(-17,0){\dashbox{1.}(7,0)}
\put(102,0){\dashbox{1.}(7,0)}
\put(-4.8,-1){$\triangle$}
%\put(-3,25){\makebox(0,0){$\beta$}}
\put(-3,-20){\dashbox{.5}(0,20)}
\put(.5,0){\circle{2}}
\put(.5,20){\makebox(0,0){$\triangle$}}
%\put(0,25){\makebox(0,0){$\gamma$}}
\put(.5,0){\dashbox{.5}(0,20)}
\put(7.5,0){\circle*{2}}
\put(7.5,-20){\makebox(0,0){$\triangle$}}
%\put(7.5,25){\makebox(0,0){$\alpha$}}
\put(7.5,-20){\dashbox{.5}(0,20)}
\put(14.5,0){\circle*{2}}
\put(14.5,20){\circle{2}}
%\put(14.5,25){\makebox(0,0){$\alpha$}}
\put(14.5,0){\dashbox{.5}(0,20)}
\put(19.5,-1){$\triangle$}
\put(21.5,-20){\circle{2}}
%\put(21.5,25){\makebox(0,0){$\beta$}}
\put(21.5,-20){\dashbox{.5}(0,20)}
\put(25,0){\circle{2}}
\put(25,20){\circle*{2}}
%\put(25,25){\makebox(0,0){$\gamma$}}
\put(25,0){\dashbox{.5}(0,20)}
\put(30.2,-1){$\triangle$}
\put(32.2,-20){\circle*{2}}
%\put(32,25){\makebox(0,0){$\beta$}}
\put(32,-20){\dashbox{.5}(0,20)}
\put(35.5,0){\circle{2}}
\put(35.5,20){\circle*{2}}
%\put(35.5,25){\makebox(0,0){$\gamma$}}
\put(35.5,0){\dashbox{.5}(0,20)}
\put(40.5,-1){$\triangle$}
\put(42.5,-20){\circle*{2}}
%\put(42.0,25){\makebox(0,0){$\beta$}}
\put(42.5,-20){\dashbox{.5}(0,20)}
\put(46,0){\circle{2}}
\put(46,20){\makebox(0,0){$\triangle$}}
%\put(46,25){\makebox(0,0){$\gamma$}}
\put(46,0){\dashbox{.5}(0,20)}
\put(53,0){\circle*{2}}
\put(53,-20){\makebox(0,0){$\triangle$}}
%\put(53,25){\makebox(0,0){$\alpha$}}
\put(53,-20){\dashbox{.5}(0,20)}
\put(60,0){\circle*{2}}
\put(60,20){\circle{2}}
%\put(60,25){\makebox(0,0){$\alpha$}}
\put(60,0){\dashbox{.5}(0,20)}
\put(65,-1){$\triangle$}
\put(67,-20){\circle{2}}
%\put(67,25){\makebox(0,0){$\beta$}}
\put(67,-20){\dashbox{.5}(0,20)}
\put(70.5,0){\circle{2}}
\put(70.5,20){\makebox(0,0){$\triangle$}}
%\put(70.5,25){\makebox(0,0){$\gamma$}}
\put(70.5,0){\dashbox{.5}(0,20)}
\put(77.5,0){\circle*{2}}
\put(77.5,-20){\makebox(0,0){$\triangle$}}
%\put(77.5,25){\makebox(0,0){$\alpha$}}
\put(77.5,-20){\dashbox{.5}(0,20)}
\put(84.5,0){\circle*{2}}
\put(84.5,20){\circle{2}}
%\put(84.5,25){\makebox(0,0){$\alpha$}}
\put(84.5,0){\dashbox{.5}(0,20)}
\put(89.5,-1){$\triangle$}
\put(91.5,-20){\circle{2}}
%\put(91.5,25){\makebox(0,0){$\beta$}}
\put(91.5,-20){\dashbox{.5}(0,20)}
\put(95,0){\circle{2}}
%\put(95,25){\makebox(0,0){$\gamma$}}
\put(95,0){\dashbox{.5}(0,20)}
\put(100.2,-1){$\triangle$}
%\put(102,25){\makebox(0,0){{\tiny 9}}}
\put(102,-20){\dashbox{.5}(0,20)}
\put(-10,0){\line(1,0){112}}
\put(-17,-2){\makebox(0,0){PD chain}}
\put(-17,20){\makebox(0,0){$\Omega$}}
\put(-10,20){\line(1,0){112}}
\put(-10,-20){\line(1,0){112}}
\put(-17,-20){\makebox(0,0){$\Gamma$}}
\end{picture}
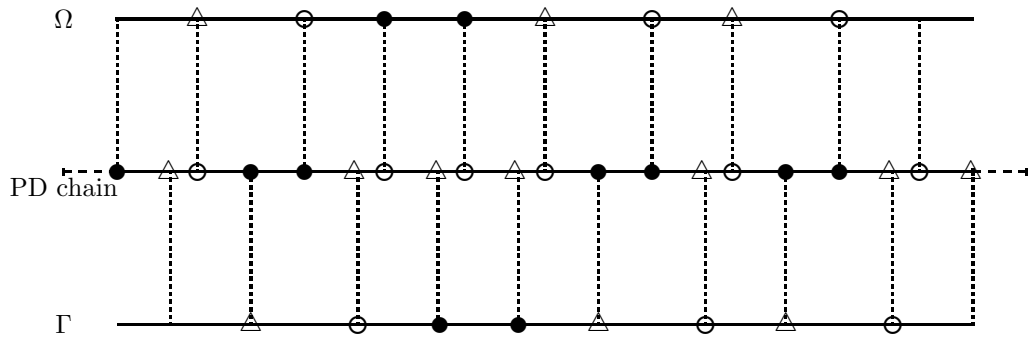
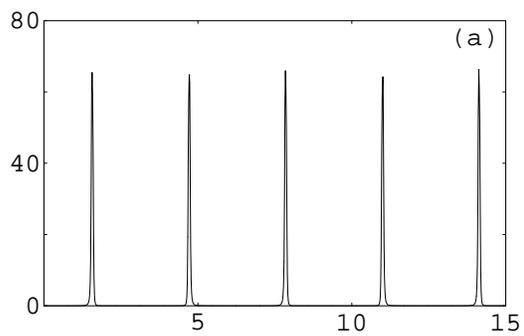
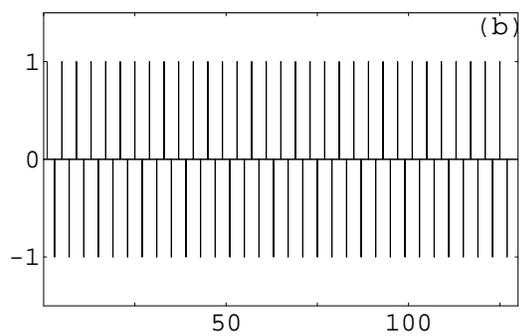
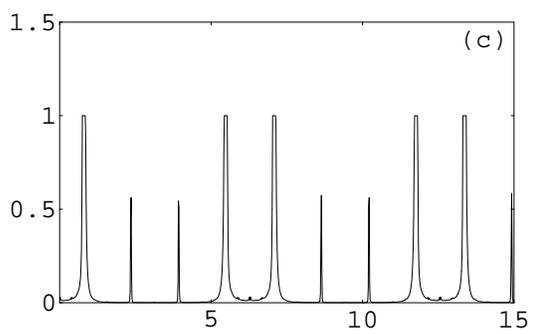
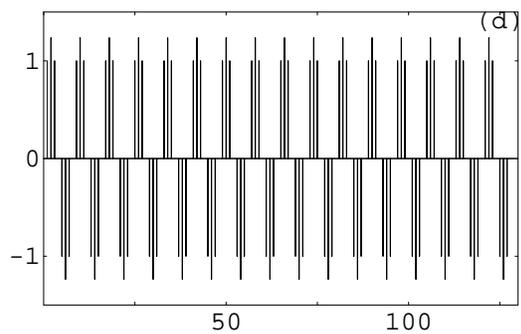
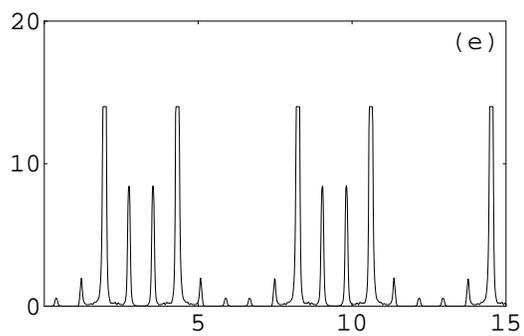
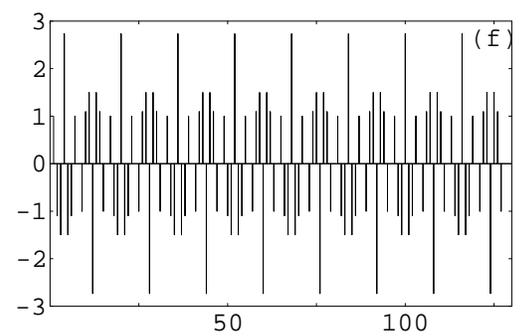
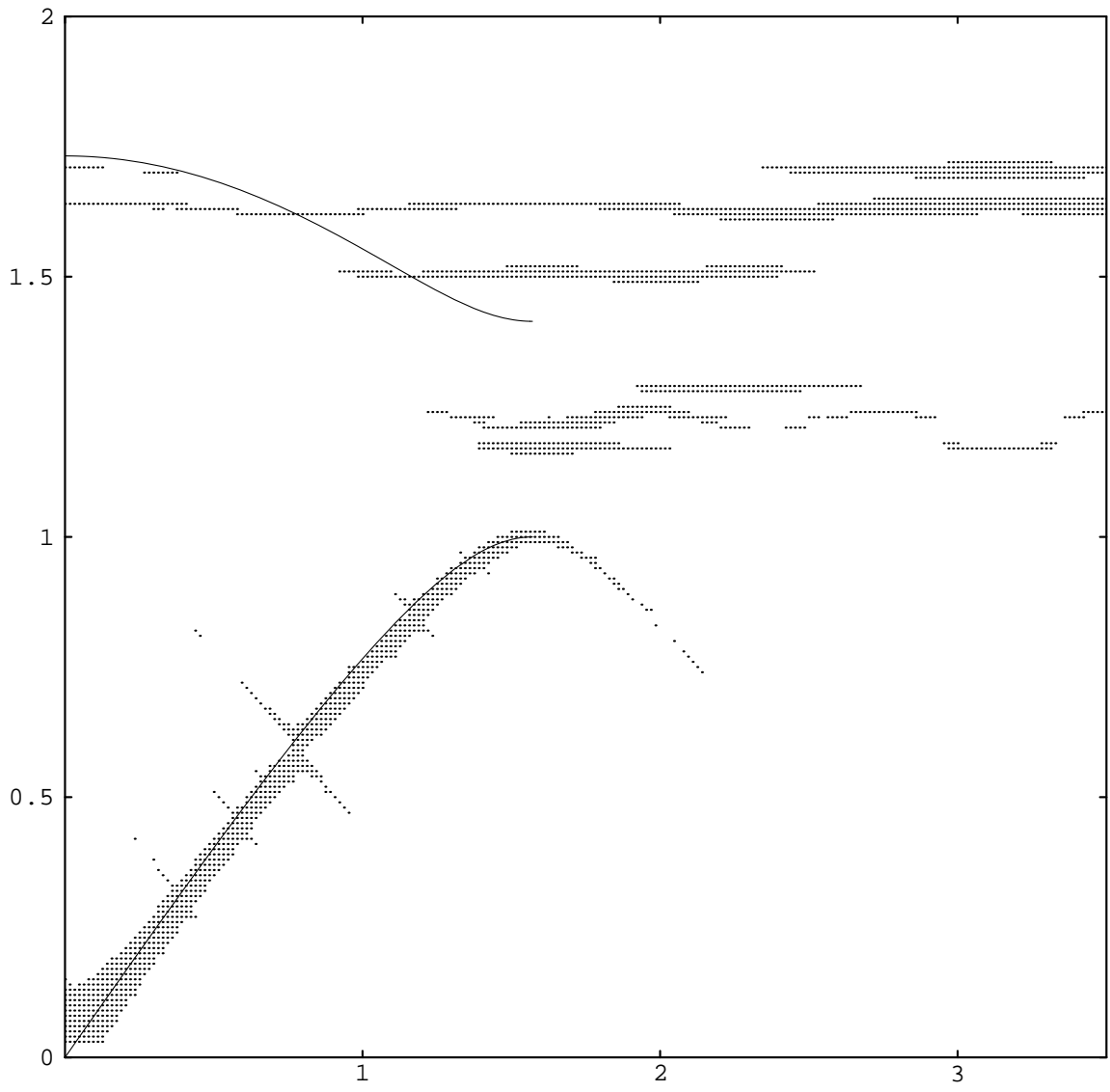
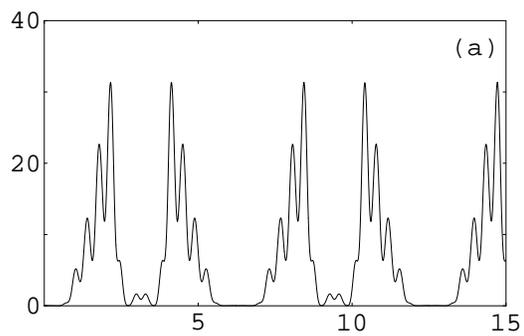
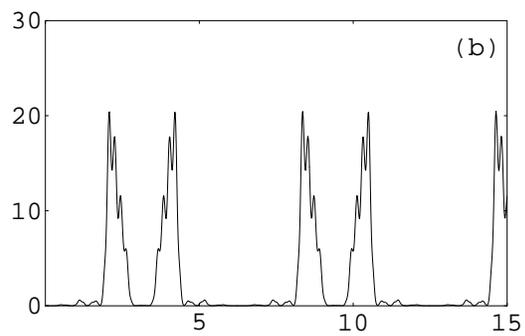
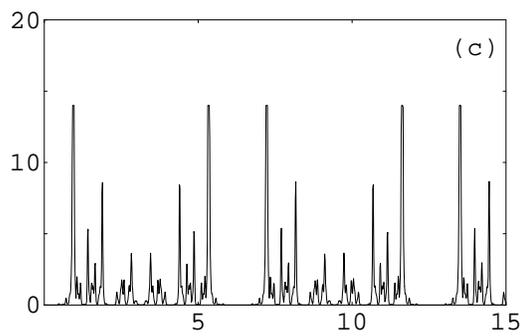
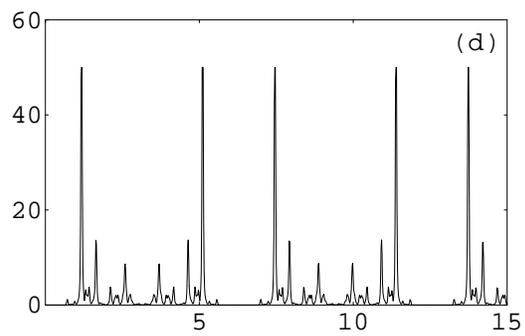
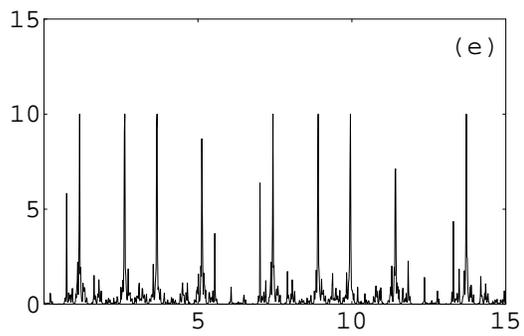
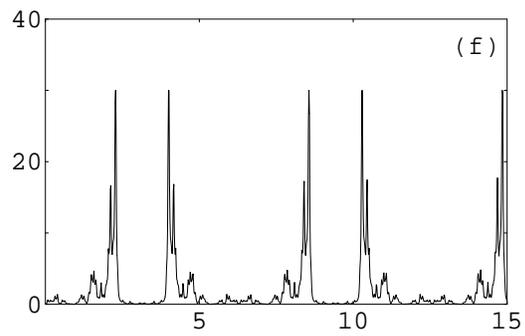
\end{document}